# Charge transport in nanoscale vertical organic semiconductor pillar devices

Janine G.E. Wilbers[1], Bojian Xu[1], Peter A. Bobbert[1,2], Michel P. de Jong[1], Wilfred G. van der Wiel[1]*.

[1]NanoElectronics Group, MESA+ Institute for Nanotechnology, University of Twente, P.O. Box 217, 7500 AE Enschede, The Netherlands

[2]Molecular Materials and Nanosystems, Department of Applied Physics, Eindhoven University of Technology, P.O. Box 513, 5600 MB Eindhoven, The Netherlands

*W.G.vanderWiel@utwente.nl.

ABSTRACT

We report charge transport measurements in nanoscale vertical pillar structures incorporating ultrathin layers of the organic semiconductor poly(3-hexylthiophene) (P3HT). P3HT layers with thickness down to 5 nm are gently top-contacted using wedging transfer, yielding highly reproducible, robust nanoscale junctions carrying high current densities (up to $10^6$ A/m$^2$). Current-voltage data modeling demonstrates excellent hole injection. This work opens up the pathway towards nanoscale, ultrashort-channel organic transistors for high-frequency and high-current-density operation.

INTRODUCTION

For application in light-emitting diodes, field-effect transistors and solar cells, organic semiconductors (OS) are playing an increasingly important role, owing to their easy processability and suitability for low-cost and flexible electronics[1,2]. Properties such as charge-carrier mobility, solution-processability, crystallinity and interface properties are important for implementation of organic semiconductors into electronic devices[3]. For many applications, like organic light-emitting transistors[4] or display pixel drivers[5], it is crucial to achieve high frequencies (~10 MHz) and large current densities (10-20 mA/cm$^2$)[6] to improve the device performance. This can be realized by choosing organic semiconductors with high carrier mobility[7,8], or by reducing the channel length down to the nanoscale[9], as demonstrated in this Letter.

For the fabrication of such nanoscale junction lengths in planar devices, source and drain electrodes have to be patterned by nanolithography techniques[10,11]. A vertical configuration, where the OS thin film is sandwiched between two (metallic) contacts, is very attractive because the channel length is defined by the thickness of the OS layer, which is very well controllable down to a few nm, whereas the device area is given by the overlap of the contacts[12,13]. Vertical geometries enable the investigation of charge transport in organic semiconductors at the nanoscale. They are already commonly used in molecular monolayer junctions[14,15], and present several advantages in comparison to planar structures. At small junction thickness, the electric field at low voltage can still be very high. The junctions can thus operate at low voltages, while maintaining sizeable current densities, which is beneficial for implementation in low-power electronic devices like organic light-emitting diodes and organic field-effect transistors[4,16].

Top contacting and nanopatterning of thin layers of OS are not straightforward. Direct metal evaporation can result in penetration through the organic film, often leading to damage of the film and electrical shorts[17,18]. Standard lithography methods that generally include optical or electron-beam lithography resists and developers, as well as lift-off procedures in solvents like acetone or dimethyl sulfoxide (DMSO), cannot be applied to pattern OS, because most of them are affected by these chemicals[19,20]. Standard lithography methods that generally include optical or electron-beam lithography resists and developers, as well as lift-off procedures in solvents like acetone or dimethyl sulfoxide (DMSO), cannot be applied to pattern OS, because most of them are affected by these chemicals. There are methods to avoid damage of the organic layer, for example, by indirect evaporation of metals onto cooled samples through a shadow mask[21] or buffer-layer-assisted deposition[22]. However, those techniques do not allow lateral nanostructuring. Other approaches for top contacting organic thin films and/or monolayers are so-called ``soft-landing'' techniques such as transfer printing[23,24], the use of conductive polymers[25] or liquid metals[26] as top contacts, lift-off float-on (LOFO)[27] and polymer-assisted lift-off (PALO)[28]. These methods, which are mainly utilized for the fabrication of molecular tunnel junctions, all have their own advantages and disadvantages. They either introduce an additional resistance or an oxide layer, or hazardous chemicals are used during processing. In Kleemann *et al.*[29] a photoresist compatible with OS was used, enabling direct patterning of the organic layers. However, in that study, the metal contact is directly evaporated onto the organic thin film, which can, as already mentioned, lead to electrical shorts, especially for thin organic films.

In this Report, we demonstrate a method to realize vertical organic devices with ultrashort junction lengths down to 5 nm. Thin layers of the p-type organic semiconductor regioregular poly(3-hexylthiophene) (P3HT) were gently contacted by wedging transfer[30], and subsequently further structured by dry etching. By applying this water-based technique[30], EBL-patterned metal electrodes (200 nm to 2 μm in diameter) are gently transferred onto P3HT. This is the only step in which EBL is applied. All the other steps are either self-aligning or achieved with standard photolithography. In earlier work, we showed that we were able to contact self-assembled monolayers (SAMs) of alkanethiols of different lengths by wedging transfer[31]. We now optimized this technique for contacting and patterning nanoscale P3HT devices. P3HT was used because it forms a smooth thin film that can be easily varied in thickness by changing the concentration and/or the spin-coating speed. We note that our water-based wedging-transfer technique could induce electron traps in the P3HT[32]. However, such traps are not expected to influence hole transport, which is the focus of our study. If necessary, water could be removed by (vacuum) annealing.

Regioregular P3HT is widely used for organic-based electronic devices due to its relatively high mobility as compared to other organic semiconductors[33-35], owing to crystallites that form via $\pi$ - $\pi$ inter-chain self-assembly[34]. However, the mobility does not only depend on regioregularity and molecular weight, but also on the solvent used. The higher the boiling point of the solvent, the slower the solvent evaporation during and after spin-coating, and the better the crystallinity of the thin film. A higher crystallinity gives better conductivities[35,36]. For this reason and the high solubility for P3HT we used bromobenzene as solvent[36]. The ionization potential of P3HT strongly depends on the average chain length, chain torsion and packing density of the chains, which change the delocalization length and screening of charges[37]. The injection barrier for charge carriers depends on the position of the ionization potential with respect to the work function of the electrode material. Au was used as electrode material, because a low injection barrier is expected for Au-P3HT interfaces[38,39]. In order to achieve good current injection from the electrode into the organic semiconductor with low contact resistance, the work function has to be aligned to either the highest occupied molecular orbital (HOMO) energy or the lowest unoccupied molecular orbital (LUMO) energy of the organic film [40].

RESULTS

**Device fabrication**

Figure 1 schematically shows the fabrication steps of the vertical metal-P3HT-metal pillars. P3HT was spin-coated onto clean Au bottom electrodes on Si/SiO$_2$ substrates (Figs. 1a and b). The thickness was varied by different P3HT concentrations in the solvent and spin-coating speeds. For wedging transfer hydrophilic surfaces are required. For the previous SAM devices[31], the molecules were only present on the metal bottom electrodes, while the rest of the substrate remained hydrophilic. This is different for the P3HT devices. P3HT was spin-coated over the whole substrate, leading to a completely hydrophobic surface. Therefore, we first removed P3HT from the unpatterned upper part of the SiO$_2$ substrate, and subsequently wedging transfer was performed (not shown in the figure)[30,31]. The metal top contacts (70 nm thick Au disks with a diameter between 200 nm and 2 μm) were wedge-transferred onto the P3HT. For wedging transfer the top contacts were embedded in a hydrophobic polymer, cellulose acetate butyrate (CAB). The cellulose polymer was prepared by dissolving the cellulose acetate butyrate (CAB) in ethyl acetate at a concentration of 30 mg/ml under stirring for 30 min. The important property of the cellulose polymer is its hydrophobicity. When the device is dipped onto water, the water penetrates in between the hydrophilic SiO$_2$ substrate and the hydrophobic CAB and thereby the polymer is lifted off. The Au disks are lifted off with the CAB polymer due to the low adhesion between Au and SiO$_2$. The Au disks were wedge-transferred onto the new substrate with the 5 μm wide bottom electrodes coated with P3HT by placing the substrate under an angle of ~45˚ on a grid holder in a beaker with Milli-Q water and slowly pumping out the water (Fig. 1c). In this way, the cellulose polymer first makes contact with the hydrophilic $SiO_2$ and then softly lands on the P3HT. In the next step, directional reactive ion beam etching was performed to form vertical pillars (Fig. 1d). The Au disks served as an etch mask for the P3HT. Next, the pillars were embedded in an electrically insulating layer of hydrogen silsesquioxane (HSQ), which is thinner on top of the pillars than on the substrate. The Au top contacts of the pillars were exposed by reactive ion beam etching, while the organic layer was still protected (Fig. 1e)[41]. Subsequently, in order to contact the pillar structures, a 100 nm thick metal layer was deposited and patterned by photolithography to form large contact pads (Fig. 1f). A detailed explanation of the fabrication process can be found in the experimental part.

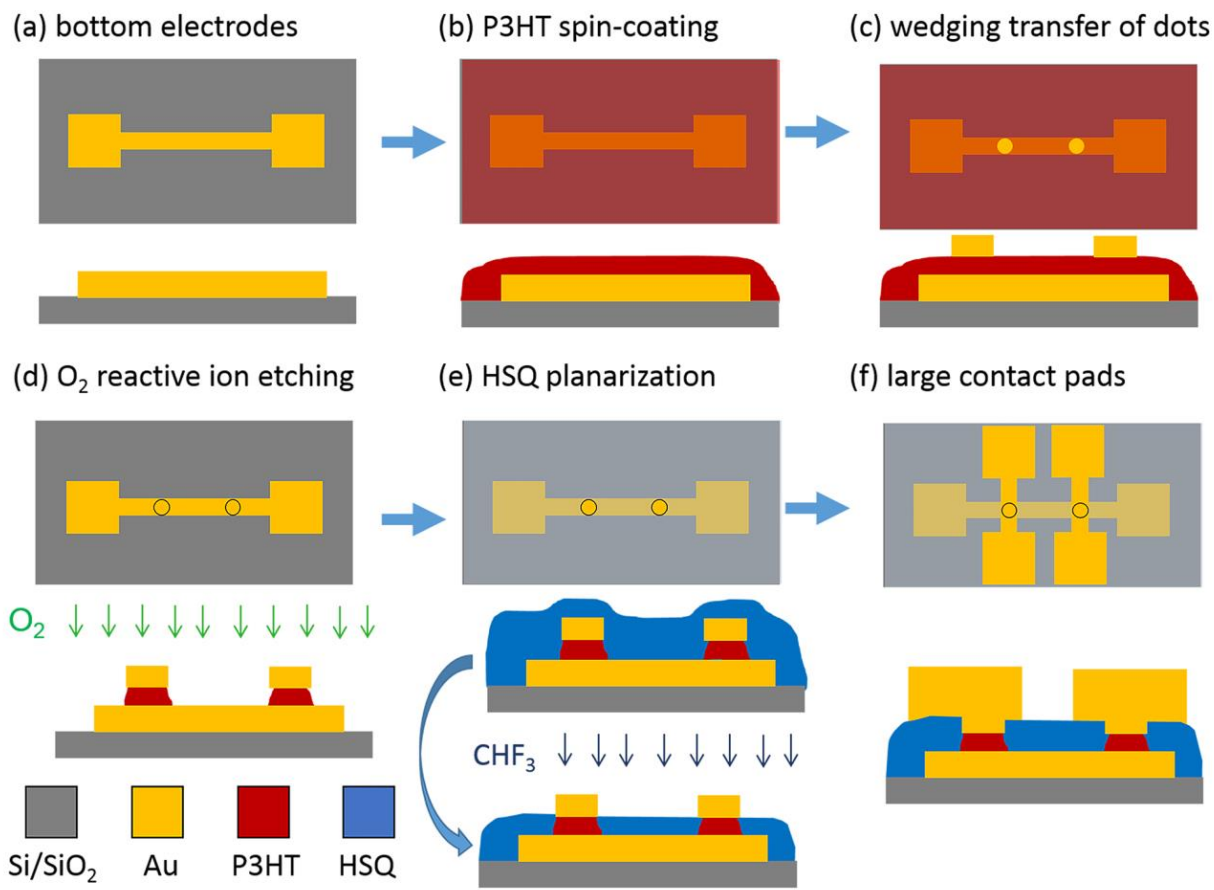

Figure 1. Fabrication steps for vertical Au-P3HT-Au pillars, showing top view (top) and side view (bottom) schematics: (a) patterning of bottom electrodes by photolithography on Si/SiO$_2$ substrates; (b) spin-coating of P3HT; (c) wedging transfer of top contacts onto the thin P3HT film[30,31]; (d) directional dry etching of vertical pillars using oxygen plasma; the top contacts served as an etch mask for the P3HT; (e) spin coating of HSQ and planarization by reactive ion etching with CHF$_3$ to open up the top Au contacts[41]; (f) evaporation of large top contacts patterned by photolithography.

Scanning electron microscopy (SEM) was used to image a cross-section of a metal-P3HT-metal test device after directional dry etching and spin coating of HSQ (see Fig. 2). The top Au served as an etch mask for the underlying P3HT thin film (see Fig. 2a). In Fig. 2c, a final test structure with large top contacts is shown.

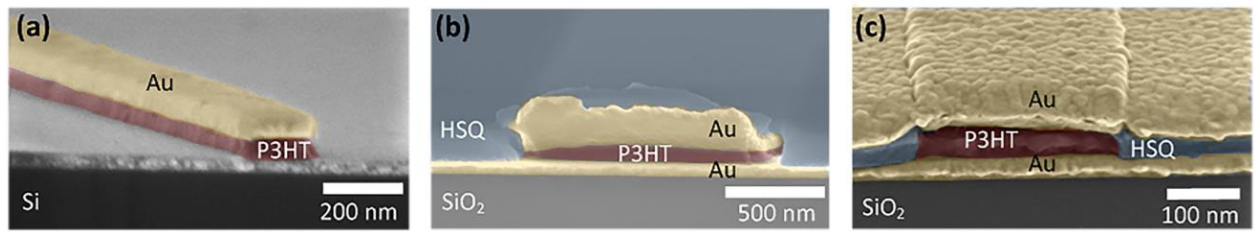

Figure 2. Scanning electron microscopy (SEM) images (false color) of a cross-section of (a) Pt-P3HT-Au test structure (200 nm wide line), (b) Au-P3HT-Au test structure (1 µm wide line) embedded in HSQ, and (c) final Au-P3HT-Au test structure (200 nm wide line) with large top contact.

**Electrical characterization**

We studied devices with four different P3HT film thicknesses, between 5 nm and 100 nm, and with different junction areas (pillar diameter between 200 nm and 2 µm). Two-terminal current-voltage (*I-V*) characterization was performed in a probe station in vacuum (< $10^{-4}$ mbar) for room-temperature (RT) measurements, and in a He cryostat for low-temperature analysis. Both setups were equipped with custom-built, low-noise electronics.

SI-Table 1 in the Supplementary Information shows an overview of all (218) measured Au-P3HT-Au junctions. The total yield of working devices is about 60%. A device is considered as "working" when it shows a non-linear *I-V* characteristic that is stable over two consecutive sweeps within a factor of 2 in the current. Junctions that were either non-conductive (n.c.), unstable (us) or shorted (s) were discarded. A non-conductive device is defined as a junction that shows current values within the noise level of the measurement equipment ($10^{-12}$ A). A device is considered unstable when it exhibits continuous current fluctuations by a factor larger than 2, far exceeding the noise.

A shorted junction has a ~100 Ω resistance or lower, and linear *I-V* characteristic. The relatively high working device yield indicates that the fabrication method is reliable, as even very thin layers in the sub-10 nm regime were successfully contacted and measured.

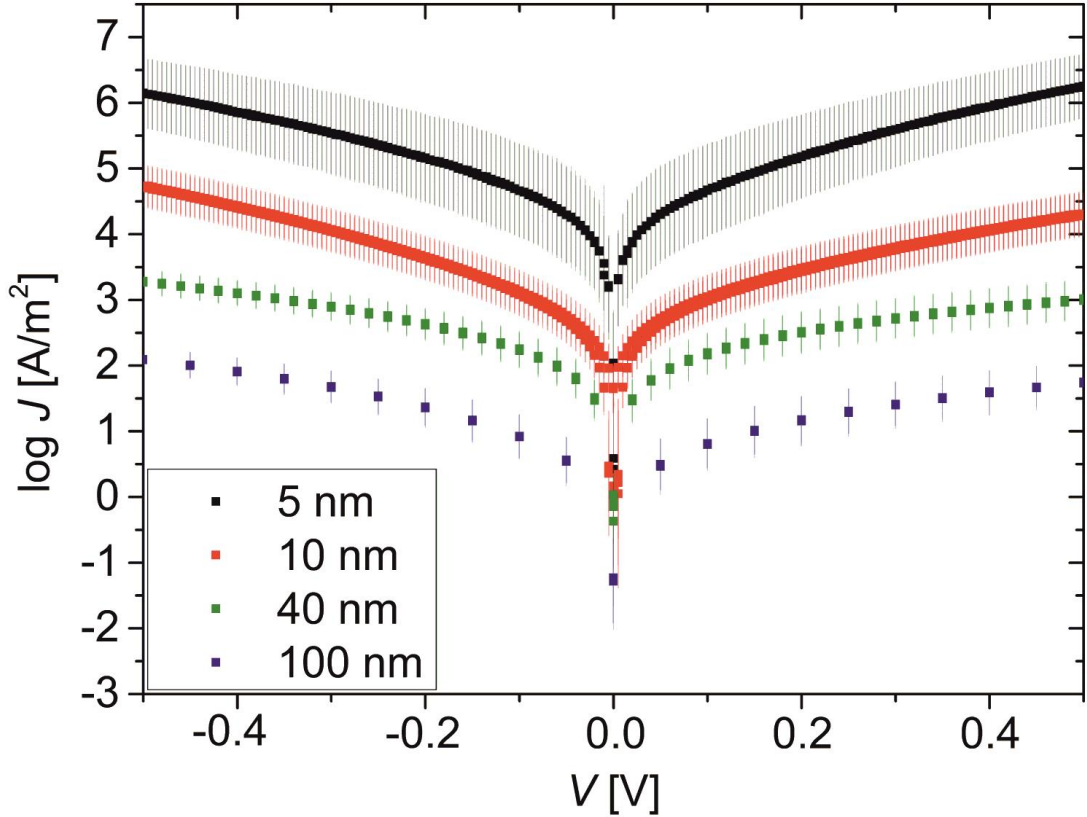

Figure 3. Electrical characterization of Au-P3HT-Au devices in a 2-point configuration at room temperature in vacuum (< $10^{-4}$ mbar); measurements were performed between ± 0.5 V, starting and ending at 0 V (black: 5 nm P3HT (steps: 5 mV), red: 10 nm P3HT (steps: 5 mV), green: 40 nm P3HT (steps: 20 mV), blue: 100 nm (steps: 50 mV)). The graph shows the logarithmic average of the current density *J* over all pillar diameters for each P3HT thickness as a function of applied voltage *V*. The data shown consist of two consecutive sweeps measured for each junction.

The current density versus voltage (*J-V*) characteristics of Au-P3HT-Au devices for different P3HT thicknesses are shown in Fig. 3. We defined *J* by dividing the absolute value of the measured current (|*I*|) by the nominal junction areas, and averaged log *J* over all junction diameters for each P3HT thickness. The error bars represent the standard deviation of the log *J* values and provide an indication of the spread in *J* for devices with the same P3HT thickness. This is a common way of plotting data for molecular junctions[26,42-44]. Hereby a normal distribution of *J* is assumed, because *J* is exponentially dependent on the thickness of the organic film, which can vary due to electrode surface defects and also roughness of the organic layer itself[45]. This approach is also appropriate for our devices with P3HT thicknesses of 5 nm and 10 nm. In order to treat all devices equally, we also applied this approach for 40 nm and 100 nm thick P3HT. Backward and forward *I-V* sweeps were performed, and no significant hysteresis was observed. The current density distribution for all working junctions of each thickness is given in SI-Figure 2. The highly reproducible current densities are on the order of $10^2$ up to $10^6$ A/m$^2$, depending on the P3HT thickness. Our devices with the thinnest (10 nm and 5 nm) P3HT layers carry exceptionally high current densities, to our knowledge not reported for P3HT devices so far. Devices with thicker P3HT layers show *J* values comparable to other P3HT-based devices sandwiched between indium tin oxide bottom electrodes and evaporated Au top contacts with similar P3HT thicknesses[46]. We observe clearly distinct values of *J* for different P3HT thicknesses. Continuous *I-V* sweeping showed that the curves were highly reproducible, and did not short after several measurements (SI-Figure 1). The junctions were also stable over a period of three weeks, while keeping the devices in vacuum conditions between the measurements. Some junctions showed only very small variations in current, while others showed a slight decrease in current over time (SI-Figure 1).

Although top and bottom contacts are made from the same metal (Au), we observe a slightly asymmetric *J-V* behavior, which we attribute to differences in the top and bottom Au-P3HT interfaces. P3HT was spin-coated onto UV/ozone-cleaned bottom electrodes and subsequently annealed, while the top electrodes were applied by wedging transfer onto the P3HT, which was exposed to air and water. The energy level alignment between the Au work function and the P3HT HOMO is influenced by the interface, and can thus change during processing[38]. Also, the local packing of the P3HT chains at either interface is likely to be different. It is known that there is a strong dependence of the charge transport[47,48] and energy level alignment[49] on the local packing. P3HT thin films contain crystalline areas, formed by $\pi - \pi$ interchain stacking, that are surrounded by an amorphous matrix[34]. The structure of the first nanometers is expected to be different from the bulk of the film[50]. This plays especially a role in vertical devices, while it is less critical for planar structures, where, e.g., in a field-effect transistor, the current flows in a thin region above the dielectric between the source and drain contacts.

Charge transport in P3HT is generally accepted to be characterized by a thermally activated hopping process[51,52,53]. This implies that the conductance increases with increasing temperature. We analyzed the *J-V* characteristics at temperatures between 150 K and 295 K with steps of 50 K. The *J-V* curves during the cooling-down and warming-up cycles show almost no differences, and exhibit excellent stability. In Fig. 4 the current density measured at different temperatures as a function of applied voltage, in the negative voltage regime (current flowing from bottom to top), is plotted on a double-logarithmic scale for four representative devices with different P3HT thicknesses.

DISCUSSION

The temperature-dependent *J-V* characteristics were simulated with the drift-diffusion model described by Van Mensfoort *et al.*[54]. The model makes use of a mobility function $\mu(T,c,F)$ depending on temperature *T*, and on the local charge-carrier concentration *c(x)* and electric field *F(x)* (*x* is the distance from the injection electrode), as calculated for thermally assisted hopping in between localized sites with a site density $N_t$ and random on-site energies taken from a Gaussian density of states (DOS) with standard deviation $\sigma$ [54]. In the absence of an injection barrier, the work function of the metal is aligned with the center of the DOS and we have *n(0) = n(L) = $N_t$/2* as boundary conditions for the solution of the drift-diffusion equation, where *L* is the P3HT thickness. In the presence of injection barriers $\varphi_1$ and $\varphi_2$ at the injecting and collecting contact, respectively, we obtain the boundary conditions *n(0)* and *n(L)* by evaluating the Gauss-Fermi integral [54].

From the modelling we derive a room temperature zero-concentration, zero-field mobility $\mu_0 = \mu(T = 294 \text{ K}, 0, 0) = 7\times10^{-5}$ cm$^2$/Vs, in accordance with other (vertical) devices[55], a site density $N_t = 1.5\times10^{26}$ m$^{-3}$ and a width of the Gaussian DOS $\sigma = 0.075$ eV. In the modeling, we concentrated on the thickest devices, for which the continuum drift-diffusion model is expected to work best, and for which the spread in the data (see Fig. 3) is the least. One can see in Figs. 4a and b that for representative devices with P3HT thicknesses of 100 nm and 40 nm an excellent modeling of the data is possible. We observe good agreement of modelled and measured *J-V* curves without introducing an injection barrier ($\varphi_1 = \varphi_2 = 0$), which indicates that the injection barrier is very low. The data shown in Fig. 4 are measured in the negative voltage regime because of a slightly lower injection barrier. Results of simulations of *J-V* curves for 40 nm and 100 nm P3HT thickness in the presence of an injection barrier at injecting and collecting contact ($\varphi_1 = \varphi_2 = \varphi$) are shown in SI-Figures 3 and 4. We conclude from these results that the injection barrier is not more than ~0.1 eV. The low injection barrier is an indication of the good quality of our devices and provides the prospect of high-current-density applications.

Figures 4c and d show *J-V* curves of representative devices with P3HT thicknesses of 10 and 5 nm, together with modeled curves, using exactly the same parameters as for the thicker devices. We observe that the overall shape of the curves and their temperature dependence are still rather well described. In particular, the much weaker temperature dependence than in the thicker devices is reproduced. This is a direct consequence of the low injection barrier: even at low temperatures a significant amount of holes diffuse from the Au contacts into the P3HT. The much higher average carrier density than in the thicker devices leads to a much weaker temperature dependence, in accordance with the theoretical prediction[56]. We checked that an increased injection barrier in the modeling increases the temperature dependence, finally leading to activated transport with the injection barrier as activation energy. The weak temperature dependence of the *J-V* curves for devices with small P3HT thickness is thus a direct indication of the good carrier injection in our devices.

However, a clear shortcoming of the modeling is that the modeled *J-V* curves are, for P3HT thicknesses of 5 and 10 nm, one to two orders of magnitude too high. We checked that such a large difference cannot be explained by an actual P3HT thickness that is higher than the nominal thickness (an unreasonably large deviation would have to be assumed) or by thickness variations in the P3HT film. The most reasonable explanation seems to be the ordering of the P3HT close to the Au electrodes. It has been reported that the first few layers of P3HT on a silicon oxide substrate often have a higher amount of edge-on orientation as compared to the bulk, which is composed of randomly dispersed π-stacked aggregates[57]. The edge-on orientation facilitates charge transport in the direction along the substrate, but severely impedes it in the vertical direction. In the case that also on Au electrodes in our devices the first few nanometers have an edge-on orientation, the conductivity of the devices with 10 nm and 5 nm P3HT thickness would be much less than in the simulations, in accordance with our finding. Edge-on orientation of P3HT at the interface with Au after an annealing procedure has indeed been reported[58,59]. For the thicker devices the effect will be much weaker, because the conductivity is then governed by the bulk.

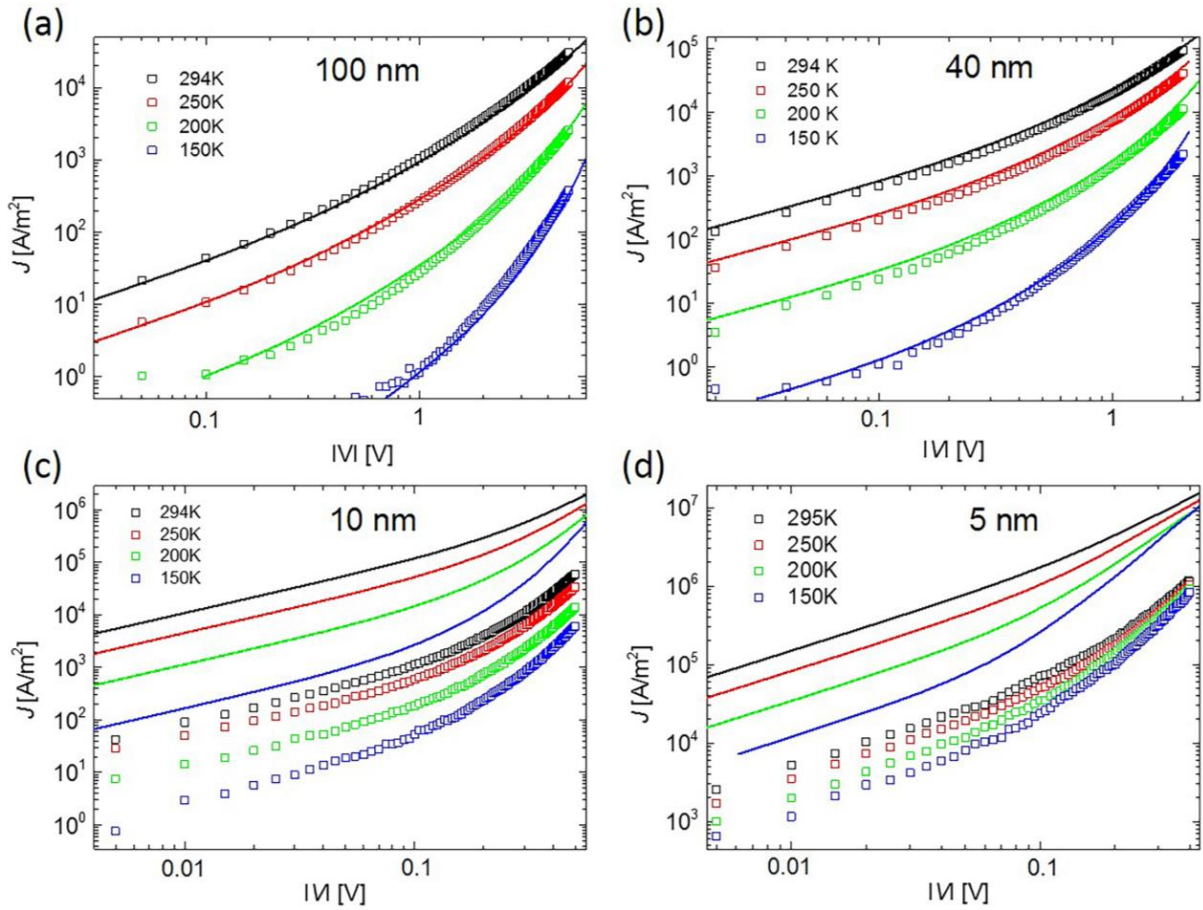

Figure 4. Experimental (symbols) and modelled (lines) current density $J$ versus applied voltage $V$ (negative voltage regime) characteristics at different temperatures for typical devices with (a) 100 nm, (b) 40 nm, (c) 10 nm and (d) 5 nm P3HT thickness and a junction diameter of 2 μm. The parameters in the drift-diffusion modeling are: room-temperature (294 K) mobility $\mu_0 = 7 \times 10^{-5}$

cm$^2$/Vs, volume density of sites $N_t = 1.5 \times 10^{26}$ m$^{-3}$, width of Gaussian DOS $\sigma$ = 75 meV. $\varepsilon_r$ =4.4 was used for the relative dielectric constant of P3HT. Barrier-less injection was assumed.

Summarizing, we fabricated sub-μm vertical Au-P3HT-Au pillars by wedging transfer, utilizing the wedge-transferred Au top contacts as etch masks for directional dry etching of the P3HT thin film. SEM images showed that the P3HT is well protected by the top contacts, and that there is a distinctive interface between the metal and the organic layer, suggesting that the metal does not penetrate the P3HT. The relatively high working device yield of 60% indicates that the top-contacting is very soft, allowing for charge transport through very thin organic films. The junctions are robust under high current densities and reveal thermally assisted hopping transport. Excellent agreement was obtained between experimental and modelled data for P3HT thicknesses of 100 nm and 40 nm. The calculated current densities for P3HT thicknesses of 10 nm and 5 nm are higher than the measured values, which we attribute to a different orientation of the chains in the P3HT thin film close to the Au electrodes as compared to the bulk. We conclude that carrier injection in our devices is very good, yielding the prospect of new types of very thin and highly conducting organic devices.

METHODS

**General information**

For all devices we used single-side polished p-type Si <100> wafers that were purchased from Okmetic. Regio-regular poly(3-hexylthiophene-2,5-diyl), cellulose acetate butyrate (average Mn ≈ 30 kDa) and bromobenzene (≥ 99.5% (GC)) were purchased from Sigma Aldrich and used as received.

**Electrode fabrication**

The wafers were cleaned for 10 minutes in nitric acid, $HNO_3$ (99%), rinsed with DI water in a quick dump rinse, followed by 10 minutes cleaning in $HNO_3$ (69%) at 95°C, quick dump rinsing and spin-drying under nitrogen flow.

Bottom electrodes were patterned on a Si wafer with 200 nm thermally grown $SiO_2$ by photolithography using an image reversal resist (Ti35ES) for metal lift-off. To this end, hexamethyldisilazane (HMDS) was spin-coated (4000 RPM) as an adhesion layer, followed by Ti35ES (4000 RPM). The photoresist was pre-baked for 120 sec at 95°C and exposed for 18 sec (EVG, EV620 Mask Aligner, Hg-lamp 12 mW/cm$^2$) through a photomask. After degassing for > 30 min the photoresist was post-baked for 120 sec at 120 °C followed by a flood exposure (60 sec, without mask). The photoresist was developed in Olin OPD 4262 (40 sec) and rinsed with DI water in a quick dump rinser. In order to remove thin resist residuals, the wafers were cleaned with UV/ozone (PR-100, UVP Inc.) for 10 min. The metal electrodes were electron-beam evaporated (BAK 600, Balzers), with a deposition rate between 0.05 and 0.2 nm/s (< $2\times10^{-6}$ mbar). For the bottom electrodes 2 nm of Ti was evaporated as an adhesion layer, followed by 20 nm of Au. The bottom substrates were cleaned with UV/ozone and rinsed with ethanol before spin-coating, to ensure clean and oxide-free electrodes.

To create the top contacts, the electron-beam resist poly(methyl methacrylate) (PMMA A4) was spin-coated at 4000 RPM for 30 minutes and baked for 3 minutes at 160 °C. Top contact structures were written on a p-type Si <100> wafer with native $SiO_2$ by electron-beam lithography (Raith150-TWO, Raith GmbH) with an aperture size of 60 μm, an acceleration voltage of 20 kV and a working distance of 10 mm. The PMMA was developed in a mixture of methyl isobutyl ketone (MIBK): isopropanol (IPA) (1:3) for 30 sec followed by 30 sec in IPA and dried under nitrogen flow. Prior to metal evaporation, the samples were treated with UV/ozone cleaning (PR-100, UVP Inc.) for 2 minutes to guarantee a clean native silicon surface without resist residuals. 70 nm of Au was evaporated to form top contacts. Metal lift-off was performed in VLSI acetone for 10 minutes with low-power sonication. Subsequently, the samples were rinsed with VLSI IPA and dried under nitrogen flow.

**Organic semiconductor preparation**

The regio-regular P3HT was dissolved in bromobenzene (2, 8, 20 mg/ml) at 80 °C for > 4 hours under stirring. After letting it cool down to room temperature under stirring, the solution was filtered through a 0.2-µm syringe filter. Before spin-coating of P3HT, the substrates were cleaned with VLSI acetone and IPA, respectively, for 10 minutes, followed by rinsing with VLSI ethanol to remove possibly present Au oxide. The P3HT solution was spin-coated for one minute at speeds between 500 RPM and 5000 RPM, leading to P3HT film thicknesses from 5 nm to 100 nm. After spin-coating, the P3HT film was annealed at 100 °C for 1 hour to let the solvent evaporate. The resulting thickness was measured by atomic force microscopy and a surface profiler.

**Wedging transfer and pillar etching**

To enable wedging transfer, the top contacts were cleaned for 10 minutes by a UV/ozone treatment to remove thin resist residues and to increase the hydrophilicity of the native silicon oxide layer. The substrate was dipped into a solution of cellulose acetate butyrate (CAB) dissolved in ethyl acetate (~30 mg/ml). The CAB layer was allowed to dry for two minutes and was subsequently removed at the edges of the substrate by dipping it into ethyl acetate. When the device was dipped under an angle of about 70° into Milli-Q water, the water penetrated at the interface between the hydrophilic $SiO_2$ substrate and the hydrophobic CAB polymer. The CAB polymer was thereby lifted off, including the Au structures, due to the low adhesion of Au to the $SiO_2$. The metal structures embedded in the CAB polymer floating at the water interface were transferred onto the new substrate with the Ti/Au bottom electrodes covered by the thin film of P3HT. To this end, the substrate was held under an angle in the water underneath the CAB with the metal top electrodes, allowing it to gently make contact with the P3HT layer. The bottom electrodes were 5 µm wide. A precise alignment of the wedge-transferred top electrodes is not needed, because the distance between the transferred Au disks is the same as the width of the bottom electrodes. This guarantees that always only one top electrode is contacted. A large area of Au disks was prepared for wedging transfer, with markers observable by the naked eye for a rough alignment with respect to the bottom electrodes. The devices were dried under ambient conditions overnight. After drying, the CAB polymer was removed using ethyl acetate. P3HT does not dissolve in this solvent and was thus not affected by this step.

The wedge-transferred top contacts were utilized as etch masks during directional dry etching by oxygen plasma (20 sccm, 100 mTorr, 10 Watt, 30-120 sec depending on the P3HT thickness) in a reactive ion etch (RIE) system. The directional dry etch enables the formation of vertical pillars, removing the P3HT everywhere except under the top contacts.

**HSQ planarization and application of large contact pads for wire bonding**

Hydrogen silsesquioxane (HSQ) (DC XR 1541-006 from Dow Corning) was spin-coated at 1000 RPM and cured for 120 sec at 120 °C resulting in a 160 nm thick film on the bottom electrodes and 80 nm on top of the pillars. Planarization of HSQ was realized by dry reactive ion etching in $CHF_3$, He, $O_2$ plasma (Adixen AMS100DE) at -10 °C and a pressure of $8\times10^{-3}$ mbar. The etch rate was 70 nm/min. When the top of the pillars was opened the etching was stopped and 100 nm Au with 2 nm Ti as an adhesion layer was evaporated (BAK 600, Balzers; deposition rate between 0.05 and 0.2 nm/s ($< 2\times10^{-6}$ mbar)). This top Au layer was photolithographically patterned. For this, HMDS was spin-coated (4000 RPM) prior to OIR 907-17 (Arch Chemicals, Inc.; 4000 RPM). The photoresist was baked for 120 sec at 95 °C before exposure. The photoresist was exposed for 4 sec (EVG, EV620 Mask Aligner, Hg-lamp 12 mW/cm$^2$) through a photomask and post-baked for 120 sec at 120 °C . It was then developed in Olin OPD 4262 (60 sec) and rinsed with DI water in a quick dump rinser. Ion beam etching (Oxford i300) with Ar ions was utilized to pattern the top Au layer.

The distance between the Au dots was chosen in such a way that only one dot was placed between the bottom electrode and the large top contacts that were applied in the last fabrication step.

**Electrical measurements**

The electrical *I-V* measurements were done in a low-temperature probe-station (Janis ST-500) connected to custom-built low-noise electronics (IVVI-DAC rack, Quantum Transport designed instrumentation, designed by Ing. Raymond Schouten from Delft University of Technology[60]) in vacuum (< 10$^{-4}$ mbar) controlled by a LabVIEW program.

The temperature-dependent transport measurements were done using a closed-cycle He refrigerator (Oxford Instruments) in a two-terminal configuration connected to the low-noise measurement electronic set-up.

**SEM and AFM imaging**

For imaging of the pillars and for thickness determination of the P3HT films, atomic force microscopy (AFM) under ambient conditions with a Veeco (Bruker) Dimension 3100 was used. Images were recorded in tapping mode using a rectangular silicon cantilever (nanosensors PPP-NCHR) with a tip diameter of ≈ 7 nm and a spring constant of 42 N/m. Furthermore, a surface profiler Bruker Dektak 8 with a 2.5 μm stylus was utilized for measurement of the P3HT thickness.

The cross-section of the pillars was imaged by high-resolution scanning electron microscopy (FEI, Sirion), FEI Focused Ion Beam System (FIB) and Zeiss Merlin HR-SEM.

ADDITIONAL INFORMATION

Supplementary Information

Further information is provided about working device yield, AFM analysis of P3HT and J-V characteristics for different junction diameters of each thickness.

Acknowledgements

This work was financially supported by the NWO-nano (STW) program, grant no. 11470 (W.G. van der Wiel) and the China Scholarship Council (grant no. 201206090154). The authors thank Martin H. Siekman, Thijs Bolhuis, Johnny G.M. Sanderink and Mark A. Smithers for technical support and SEM imaging. We thank Prof. dr. Reinder Coehoorn for providing the computer code to model the *J-V* curves.

Author Contributions

W.G. van der Wiel conceived and supervised the project. Device fabrication and writing of the manuscript was done by J.G.E. Wilbers. Electrical measurements were performed by B. Xu and J.G.E. Wilbers. Simulations were done by P.A. Bobbert and B. Xu. W.G. van der Wiel, P.A. Bobbert, M.P. de Jong, B. Xu and J.G.E. Wilbers discussed and analyzed the results. All authors have given approval to the final version of the manuscript.

Competing financial interests

The authors declare no competing financial interests.

# Supplementary information: Charge transport in nanoscale vertical organic semiconductor pillar devices

Janine G.E. Wilbers[1], Bojian Xu[1], Peter A. Bobbert[1,2], Michel P. de Jong[1], Wilfred G. van der Wiel[1]*.

[1]NanoElectronics Group, MESA+ Institute for Nanotechnology, University of Twente, P.O. Box 217, 7500 AE Enschede, The Netherlands

[2]Molecular Materials and Nanosystems, Department of Applied Physics, Eindhoven University of Technology, P.O. Box 513, 5600 MB Eindhoven, The Netherlands

*W.G.vanderWiel@utwente.nl.

**SI-Table 1.** Overview of working Au-P3HT-Au devices for different P3HT thicknesses and pillar diameters (s: shorted device; n.c.: non-conductive device; us: unstable device).

| P3HT thickness [nm] | Pillar diameter [µm] | Total number of junctions | working junctions | non-working junctions |
|---|---|---|---|---|
| **5** | 2 | 9 | 1 | 7 (s), 1 (us) |
| | 1 | 11 | 6 | 3 (s), 1 (n.c.) 1 (us) |
| | 0.5 | 16 | 8 | 6 (s), 2 (n.c.) |
| | 0.4 | 1 | 1 | |
| | 0.3 | 15 | 4 | 10 (s), 1 (n.c.) |
| | 0.2 | 1 | 1 | |
| **10** | 2 | 29 | 19 | 8 (s), 2 (n.c.) |
| | 1 | 35 | 22 | 10 (s), 3 (n.c.) |
| | 0.5 | 39 | 26 | 1, (s), 12 (n.c.) |
| | 0.2 | 12 | | 10 (n.c.), 2 unstable |
| **40** | 2 | 8 | 7 | 1 (s) |
| | 1 | 10 | 9 | 1 (n.c.) |
| | 0.5 | 9 | 7 | 1 (s), 1 (n.c.) |
| | 0.2 | 9 | 6 | 3 (n.c.) |
| **100** | 2 | 4 | 4 | |
| | 1 | 9 | 7 | 2 (n.c.) |
| **total** | | **217** | **128** | **89** |

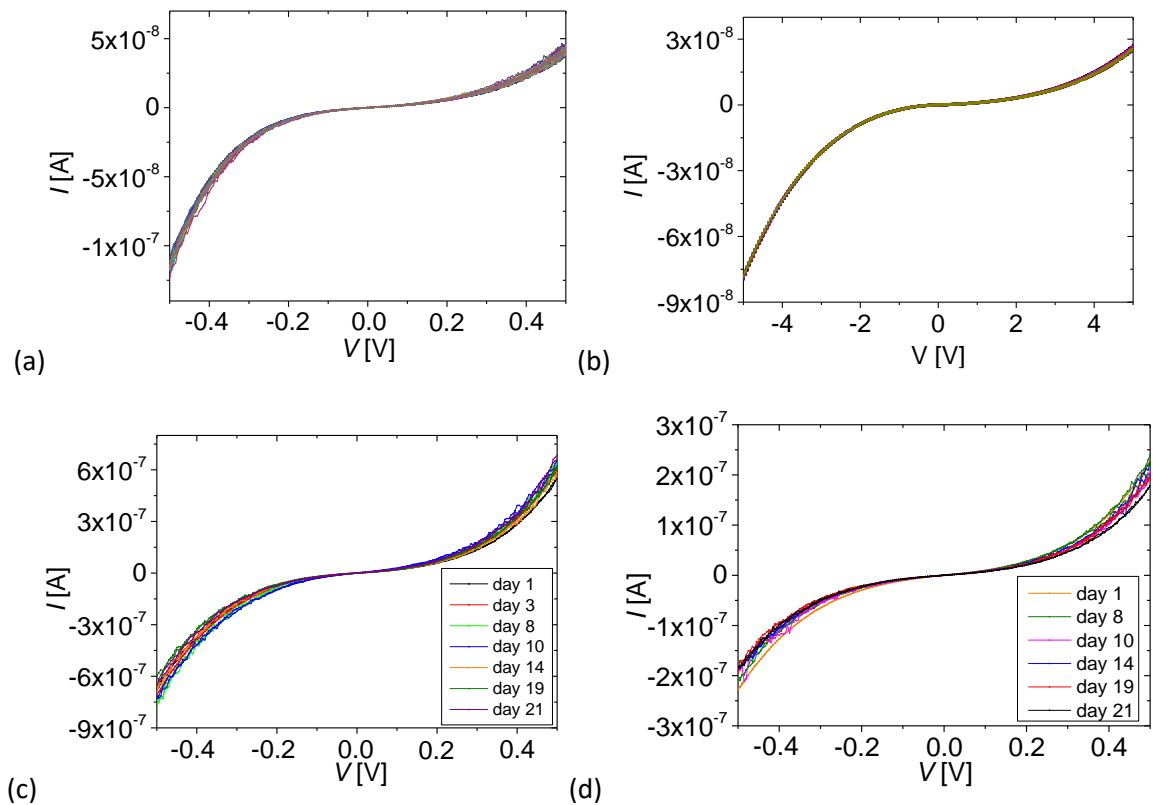

**SI-Figure 1.** Stability of the Au-P3HT-Au devices measured at room temperature in vacuum (< $10^{-4}$ mbar) between ± 0.5 V with steps of 0.005 V (a, c, d) and ± 5 V with steps of 0.05 V (b). (a) 40 consecutive I-V sweeps of a device with 10 nm P3HT and a junction diameter of 1 μm; (b) 10 consecutive I-V sweeps of a device with 100 nm P3HT and a junction diameter of 2 μm. I-V sweeps over a time window of three weeks for a device with (c) 5 nm P3HT and a junction diameter of 500 nm and (d) 5 nm P3HT and a junction diameter of 1 μm, respectively.

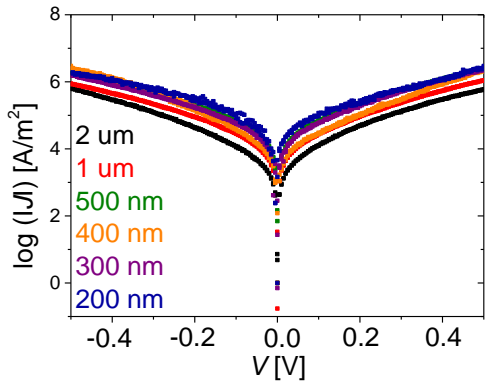
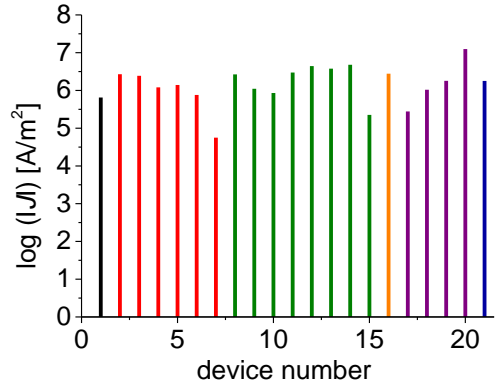

(a)

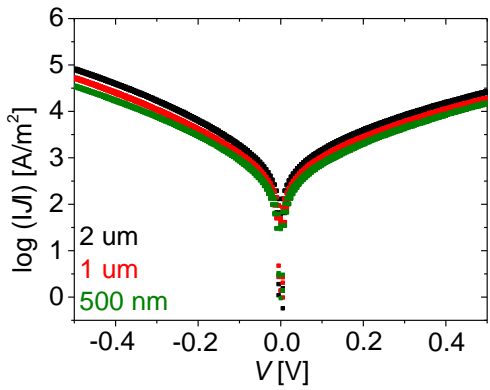
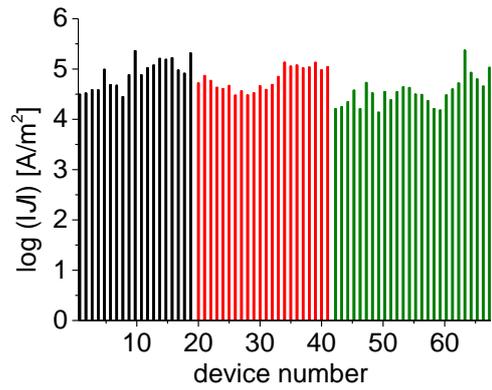

(b)

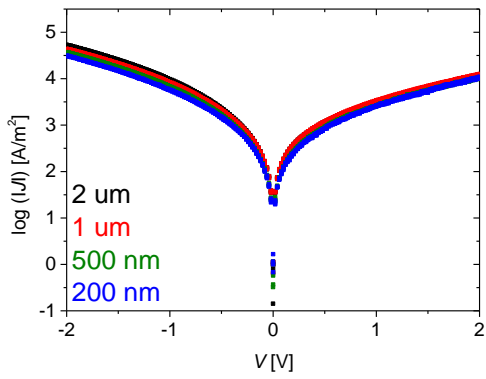
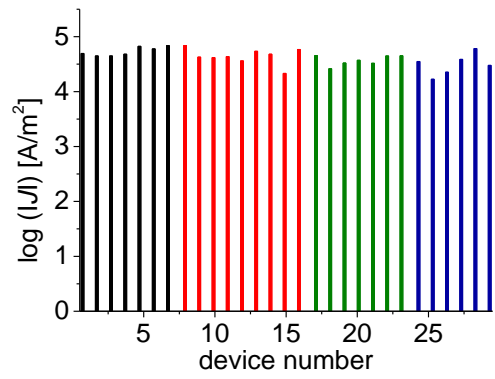

(c)

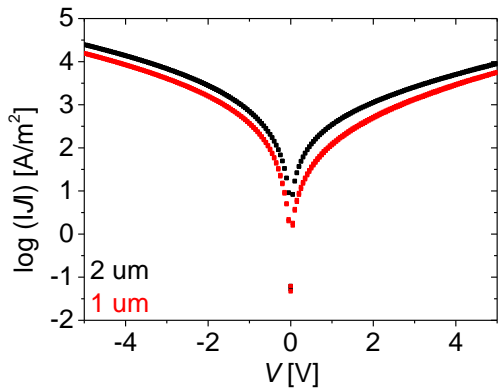
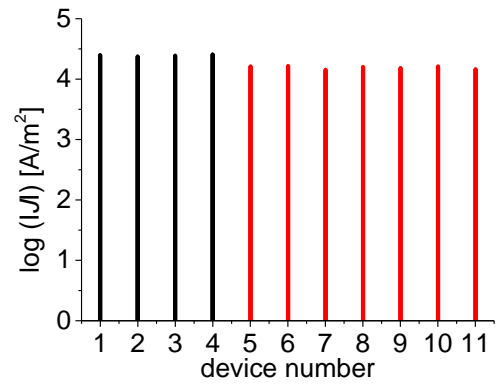

(d)

**SI-Figure 2.** Electrical characterization of Au-P3HT-Au devices in 2-point configuration at room temperature in vacuum (< $10^{-4}$ mbar) between ± 0.5 V with steps of 0.005 V for (a) 5 nm P3HT and (b) 10 nm, (c) 40 nm P3HT between ± 2 V with steps of 0.02 V and (d) 100 nm P3HT between ± 5 V with steps of 0.05 V. The curves were obtained by taking the logarithmic average of the current densities measured in devices with the same indicated junction diameter. The graphs on the right side show the corresponding distribution of log (|J|) at -0.5 V.

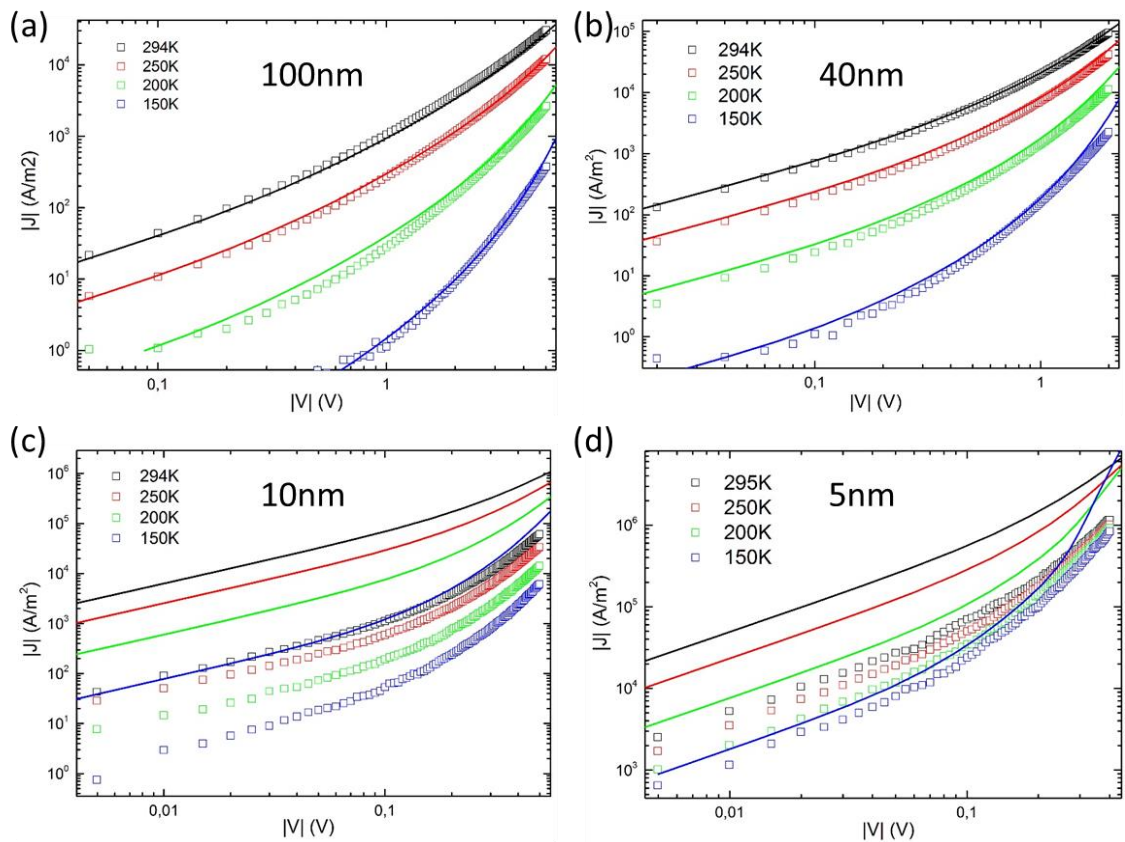

**SI-Figure 3.** Experimental (symbols) and modelled (lines) current density *J* versus applied voltage *V* (negative voltage regime) characteristics of a junction with 2 µm diameter and (a) 100 nm, (b) 40 nm, (c) 10 nm and (d) 5 nm P3HT thickness at various temperatures with an injection barrier of 0.1 eV. The parameters in the drift-diffusion modeling are: room-temperature (294 K) mobility $\mu_0 = 7.3 \times 10^{-9}$ m$^2$/Vs, volume density of molecules $N_t = 1.2 \times 10^{26}$ m$^{-3}$, width of Gaussian DOS $\sigma = 72.5$ meV. The relative dielectric constant used for P3HT is $\varepsilon_r = 4.4$.

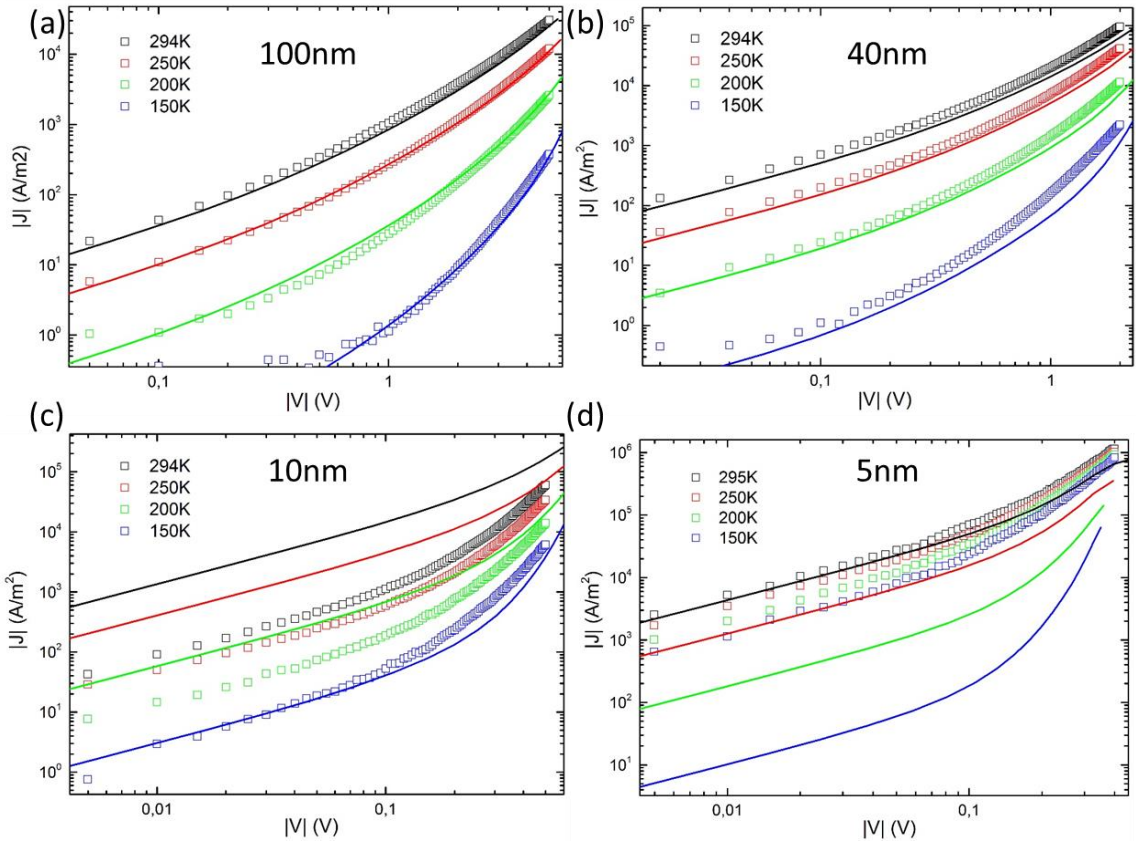

**SI-Figure 4.** The same as SI-Figure 3, but for an injection barrier of 0.2 eV. The parameters in the drift-diffusion modeling are: $\mu_0 = 8\times10^{-9}$ m$^2$/Vs, $N_t = 1.0\times10^{26}$ m$^{-3}$, $\sigma = 71.25$ meV.